\documentclass[english,11pt, a4paper]{article}
\usepackage[latin9]{inputenc}
\usepackage{geometry}
\usepackage{color}
\usepackage{babel}
\usepackage{amsmath}
\usepackage{amsthm}
\usepackage{amssymb}
\usepackage{mleftright}
\usepackage[unicode=true,pdfusetitle,
 bookmarks=true,bookmarksnumbered=false,bookmarksopen=false,
 breaklinks=false,pdfborder={0 0 0},pdfborderstyle={},backref=false,colorlinks=true]
 {hyperref}
\hypersetup{
 linkcolor=vblue, citecolor=vblue}

\makeatletter

\DeclareTextSymbolDefault{\textquotedbl}{T1}

\numberwithin{equation}{section}
\theoremstyle{plain}
\newtheorem{thm}{\protect\theoremname}
\theoremstyle{plain}
\newtheorem*{thm*}{\protect\theoremname}
\theoremstyle{plain}
\newtheorem{prop}[thm]{\protect\propositionname}

\definecolor{vblue}{RGB}{0,0,255}

\allowdisplaybreaks

\makeatother

\providecommand{\propositionname}{Proposition}
\providecommand{\theoremname}{Theorem}

\begin{document}
\title{Hilbert-Schmidt Estimates for Fermionic 2-Body Operators}
\author{Martin Ravn Christiansen\\
\\
{\footnotesize{}Department of Mathematics, Ludwig Maximilian University
of Munich, Germany}\\
{\footnotesize{}Email: martin.christiansen@math.lmu.de}}
\maketitle
\begin{abstract}
We prove that the 2-body operator $\gamma_{2}^{\Psi}$ of a fermionic
$N$-particle state $\Psi$ obeys $\Vert\gamma_{2}^{\Psi}\Vert_{\mathrm{HS}}\leq\sqrt{5}N$,
which complements the bound of Yang\cite{Yang-62} that $\Vert\gamma_{2}^{\Psi}\Vert_{\mathrm{op}}\leq N$.
This estimate furthermore resolves a conjecture of Carlen-Lieb-Reuvers\cite{CarLieReu-16} concerning the entropy of the normalized $2$-body
operator.

We also prove that the Hilbert-Schmidt norm of the truncated $2$-body
operator $\gamma_{2}^{\Psi,T}$
obeys the inequality $\Vert\gamma_{2}^{\Psi,T}\Vert_{\mathrm{HS}}\leq\sqrt{5N\,\mathrm{tr}\,(\gamma_{1}^{\Psi}(1-\gamma_{1}^{\Psi}))}$.
\end{abstract}

\section{Introduction}

Let $\mleft(\mathfrak{h},\left\langle \cdot,\cdot\right\rangle \mright)$
be a (separable) Hilbert space and consider the space of fermionic
$N$-particle states, $\bigwedge^{N}\mathfrak{h}$. Given a normalized
state $\Psi\in\bigwedge^{N}\mathfrak{h}$, one defines \textit{the
1- and 2-body operators associated to $\Psi$}, $\gamma_{1}^{\Psi}:\mathfrak{h}\rightarrow\mathfrak{h}$
and $\gamma_{2}^{\Psi}:\mathfrak{h}\otimes\mathfrak{h}\rightarrow\mathfrak{h}\otimes\mathfrak{h}$,
by
\begin{equation}
\left\langle \varphi_{1},\gamma_{1}^{\Psi}\psi_{1}\right\rangle =\left\langle \Psi,c^{\ast}\mleft(\psi_{1}\mright)c\mleft(\varphi_{1}\mright)\Psi\right\rangle 
\end{equation}
and
\begin{equation}
\left\langle \mleft(\varphi_{1}\otimes\varphi_{2}\mright),\gamma_{2}^{\Psi}\mleft(\psi_{1}\otimes\psi_{2}\mright)\right\rangle =\left\langle \Psi,c^{\ast}\mleft(\psi_{1}\mright)c^{\ast}\mleft(\psi_{2}\mright)c\mleft(\varphi_{2}\mright)c\mleft(\varphi_{1}\mright)\Psi\right\rangle 
\end{equation}
for any $\varphi_{1},\varphi_{2},\psi_{1},\psi_{2}\in\mathfrak{h}$.
Here $c^{\ast}\mleft(\cdot\mright)$ and $c\mleft(\cdot\mright)$ denote
the fermionic creation and annihilation operators, which obey the
canonical anticommutation relations (CAR)
\begin{equation}
\left\{ c\mleft(\varphi\mright),c^{\ast}\mleft(\psi\mright)\right\} =\left\langle \varphi,\psi\right\rangle ,\quad\left\{ c\mleft(\varphi\mright),c\mleft(\psi\mright)\right\} =0=\left\{ c^{\ast}\mleft(\varphi\mright),c^{\ast}\mleft(\psi\mright)\right\} ,\quad\varphi,\psi\in\mathfrak{h}.
\end{equation}
We recall some well-known properties of $\gamma_{1}^{\Psi}$ and $\gamma_{2}^{\Psi}$:
Firstly, they are non-negative. This is obvious in the case of $\gamma_{1}^{\Psi}$,
and for $\gamma_{2}^{\Psi}$ this follows by noting that if $\mleft(u_{k}\mright)_{k=1}^{\infty}$
is an orthonormal basis for $\mathfrak{h}$, then a general tensor
$\Phi\in\mathfrak{h}\otimes\mathfrak{h}$ can be written as
\begin{equation}
\Phi=\sum_{k,l=1}^{\infty}\overline{\Phi_{k,l}}\mleft(u_{k}\otimes u_{l}\mright)
\end{equation}
for $\Phi_{k,l}=\left\langle \Phi,u_{k}\otimes u_{l}\right\rangle $,
from which it readily follows that the inner product $\left\langle \Phi,\gamma_{2}^{\Psi}\Phi\right\rangle $
is given by
\begin{equation}
\left\langle \Phi,\gamma_{2}^{\Psi}\Phi\right\rangle =\left\Vert \sum_{k,l=1}^{\infty}\Phi_{k,l}c_{l}c_{k}\Psi\right\Vert ^{2}\label{eq:ActiononGeneralTensor}
\end{equation}
where $c_{k}^{\ast}=c^{\ast}\mleft(u_{k}\mright)$ and $c_{k}=c\mleft(u_{k}\mright)$
denotes the creation and annihilation operators associated to the
basis $\mleft(u_{k}\mright)_{k=1}^{\infty}$.

Secondly, they are trace-class, with
\begin{equation}
\mathrm{tr}\mleft(\gamma_{1}^{\Psi}\mright)=N,\quad\mathrm{tr}\mleft(\gamma_{2}^{\Psi}\mright)=N\mleft(N-1\mright),
\end{equation}
which follows from the identity $\sum_{k=1}^{\infty}c_{k}^{\ast}c_{k}=M\,\mathrm{id}_{\bigwedge^{M}\mathfrak{h}}$,
which holds on $\bigwedge^{M}\mathfrak{h}$ for any $M\in\mathbb{N}$.

This combined with their non-negativity implies the trivial estimates
$\left\Vert \gamma_{1}^{\Psi}\right\Vert _{\mathrm{op}}\leq N$ and
$\left\Vert \gamma_{2}^{\Psi}\right\Vert _{\mathrm{op}}\leq N\mleft(N-1\mright)$.
In the \textit{bosonic} case these are in fact optimal, but this is
not so for the fermionic case: For $\gamma_{1}^{\Psi}$ one may note
that
\begin{equation}
\left\langle \varphi,\gamma_{1}^{\Psi}\varphi\right\rangle =\left\langle \Psi,c^{\ast}\mleft(\varphi\mright)c\mleft(\varphi\mright)\Psi\right\rangle \leq\left\langle \Psi,\left\{ c^{\ast}\mleft(\varphi\mright),c\mleft(\varphi\mright)\right\} \Psi\right\rangle =\left\langle \varphi,\varphi\right\rangle ,\quad\varphi\in\mathfrak{h},\label{eq:StandardFermionicEstimate}
\end{equation}
where we simply added the non-negative quantity $\left\Vert c^{\ast}\mleft(\varphi\mright)\Psi\right\Vert ^{2}$
and applied the CAR. This shows that $\left\Vert \gamma_{1}^{\Psi}\right\Vert _{\mathrm{op}}\leq1$,
which is usually described as a consequence of the \textit{Fermi exclusion
principle}, which informally speaking prohibits more than one fermion
from occupying any given one-particle state.

In terms of the operators $c_{k}$ this bound can equivalently be
phrased as
\begin{equation}
\left\Vert \sum_{k=1}^{\infty}\alpha_{k}c_{k}\right\Vert _{\mathrm{op}}\leq\sqrt{\sum_{k=1}^{\infty}\left|\alpha_{k}\right|^{2}}
\end{equation}
which we can also use to improve our bound on $\left\Vert \gamma_{2}^{\Psi}\right\Vert _{\mathrm{op}}$
as follows: By the triangle and Cauchy-Schwarz inequalities, we can
apply this to equation (\ref{eq:ActiononGeneralTensor}) for
\begin{equation}
\sqrt{\left\langle \Phi,\gamma_{2}^{\Psi}\Phi\right\rangle }\leq\sum_{k=1}^{\infty}\left\Vert \mleft(\sum_{l=1}^{\infty}\Phi_{k,l}c_{l}\mright)c_{k}\Psi\right\Vert \leq\sqrt{\sum_{k,l=1}^{\infty}\left|\Phi_{k,l}\right|^{2}}\sqrt{\sum_{k=1}^{\infty}\left\Vert c_{k}\Psi\right\Vert ^{2}}=\sqrt{N}\left\Vert \Phi\right\Vert \label{eq:2-BodyOperatorEstimate}
\end{equation}
which implies that $\left\Vert \gamma_{2}^{\Psi}\right\Vert _{\mathrm{op}}\leq N$
- a bound which Yang proved was optimal\footnote{In fact he proved the stronger statement that $\left\Vert \gamma_{2}^{\Psi}\right\Vert _{\mathrm{op}}\leq\frac{M-N+2}{M}N$
when $M=\dim\mleft(\mathfrak{h}\mright)<\infty$ and $M$ and $N$ are
even, and characterized the optimizers in this case.} in the classic paper \cite{Yang-62}.

\subsection{Main Results}

In this paper we present two results on $\gamma_{2}^{\Psi}$. The
first result concerns $\left\Vert \gamma_{2}^{\Psi}\right\Vert _{\mathrm{HS}}$:
Given the identity $\left\Vert \gamma_{2}^{\Psi}\right\Vert _{\mathrm{tr}}=\mathrm{tr}\mleft(\gamma_{2}^{\Psi}\mright)=N\mleft(N-1\mright)$
and the general inequality $\left\Vert \gamma_{2}^{\Psi}\right\Vert _{\mathrm{op}}\leq N$,
one can deduce that the Hilbert-Schmidt norm of $\gamma_{2}^{\Psi}$
must obey
\begin{equation}
\left\Vert \gamma_{2}^{\Psi}\right\Vert _{\mathrm{HS}}\leq\sqrt{\left\Vert \gamma_{2}^{\Psi}\right\Vert _{\mathrm{tr}}\left\Vert \gamma_{2}^{\Psi}\right\Vert _{\mathrm{op}}}\leq N\sqrt{N-1}=O\mleft(N^{\frac{3}{2}}\mright),
\end{equation}
and since $\left\Vert \gamma_{2}^{\Psi}\right\Vert _{\mathrm{tr}}=N\mleft(N-1\mright)$
is an identity and $\left\Vert \gamma_{2}^{\Psi}\right\Vert _{\mathrm{op}}\leq N$
is optimal, one might suppose that this is at least nearly optimal.
Below we will however prove that this is far from the case, as the
following holds:
\begin{thm}
\label{them:HilbertSchmidtBound}For any $N\in\mathbb{N}$ and normalized
$\Psi\in\bigwedge^{N}\mathfrak{h}$ it holds that
\[
\left\Vert \gamma_{2}^{\Psi}\right\Vert _{\mathrm{HS}}\leq\sqrt{5}N.
\]
\end{thm}

This result is interesting in two respects: Firstly, it is of the
same order (with respect to $N$) as the optimal bound $\left\Vert \gamma_{2}^{\Psi}\right\Vert _{\mathrm{op}}\leq N$,
even though $\left\Vert \gamma_{2}^{\Psi}\right\Vert _{\mathrm{tr}}=N\mleft(N-1\mright)$.
This informally implies that although it is possible for $\gamma_{2}^{\Psi}$
to have eigenvalues of order $N$, it can not have ``too many\textquotedbl{}
large eigenvalues.\smallskip{}

Secondly, it is easy to compute that for a Slater state $\Psi=u_{1}\wedge\cdots\wedge u_{N}$,
$\left\Vert \gamma_{2}^{\Psi}\right\Vert _{\mathrm{HS}}=\sqrt{2N\mleft(N-1\mright)}$,
which is also $O\mleft(N\mright)$\@. This suggests that an $O\mleft(N\mright)$
behavior of the Hilbert-Schmidt norm of fermionic 2-body operators
might be a general feature.\smallskip{}

This bound also resolves a conjecture of Carlen-Lieb-Reuvers concerning
the entropy of the trace-normalized 2-body operator $\overline{\gamma}_{2}=\frac{1}{N\mleft(N-1\mright)}\gamma_{2}^{\Psi}$.
In \cite{CarLieReu-16} they conjectured (Conjecture 2.6) that the
entropy of this should always obey a bound of the form
\begin{equation}
S\mleft(\overline{\gamma}_{2}\mright):=-\mathrm{tr}\mleft(\overline{\gamma}_{2}\log\mleft(\overline{\gamma}_{2}\mright)\mright)\geq2\log\mleft(N\mright)+O\mleft(1\mright).
\end{equation}
As noted in \cite{CarLieReu-16}, Jensen's inequality for the convex
function $x\mapsto-\log\mleft(x\mright)$ implies that
\begin{equation}
S\mleft(\overline{\gamma}\mright)\geq-\log\mleft(\mathrm{tr}\mleft(\overline{\gamma}^{2}\mright)\mright)=-\log\mleft(\left\Vert \overline{\gamma}\right\Vert _{\mathrm{HS}}^{2}\mright)
\end{equation}
for any density operator $\overline{\gamma}$, so as a consequence
of Theorem \ref{them:HilbertSchmidtBound} we find that
\begin{equation}
S\mleft(\overline{\gamma}_{2}\mright)\geq-\log\mleft(\frac{5N^{2}}{\mleft(N\mleft(N-1\mright)\mright)^{2}}\mright)=2\log\mleft(N\mright)-\log\mleft(5\mleft(1+\frac{2N-1}{\mleft(N-1\mright)^{2}}\mright)\mright)
\end{equation}
for any $N\geq2$, which is to say $S\mleft(\overline{\gamma}_{2}\mright)\geq2\log\mleft(N\mright)-\log\mleft(5\mright)+o\mleft(1\mright)$
as $N\rightarrow\infty$.

\subsubsection*{Truncated $2$-Body Operators}

Our second result concerns the \textit{truncated} $2$-body operator:
It is well-known that $\Psi$ is a Slater state if and only if $\gamma_{1}^{\Psi}$
is a projection, i.e. $\mleft(\gamma_{1}^{\Psi}\mright)^{2}=\gamma_{1}^{\Psi}$,
and that in this case the $2$-body operator can be expressed in terms
of $\gamma_{1}^{\Psi}$ as
\begin{equation}
\gamma_{2}^{\Psi}=\mleft(1-\mathrm{Ex}\mright) \, \gamma_{1}^{\Psi}\otimes\gamma_{1}^{\Psi}
\end{equation}
where $\mathrm{Ex}:\mathfrak{h}\otimes\mathfrak{h}\rightarrow\mathfrak{h}\otimes\mathfrak{h}$
acts according to $\mathrm{Ex}\mleft(\varphi\otimes\psi\mright)=\psi\otimes\varphi$.\smallskip{}

For a general state $\Psi$ one would expect that if $\Psi$ is ``nearly
Slater'', in an appropriate sense, then the same expression should
be approximately valid for $\gamma_{2}^{\Psi}$ - this leads to the
Hartree-Fock functional as an approximation to the energy of a fermionic
system, for instance.\smallskip{}

A way to make this precise is to consider the truncated $2$-body
operator $\gamma_{2}^{\Psi,T}:\mathfrak{h}\otimes\mathfrak{h}\rightarrow\mathfrak{h}\otimes\mathfrak{h}$,
defined by
\begin{equation}
\gamma_{2}^{\Psi,T}=\gamma_{2}^{\Psi}-\mleft(1-\mathrm{Ex}\mright) \, \gamma_{1}^{\Psi}\otimes\gamma_{1}^{\Psi},
\end{equation}
and establish that this can be controlled in terms of $\mathrm{tr}\mleft(\gamma_{1}^{\Psi}\mleft(1-\gamma_{1}^{\Psi}\mright)\mright)=\mathrm{tr}\mleft(\gamma_{1}^{\Psi}-\mleft(\gamma_{1}^{\Psi}\mright)^{2}\mright)$,
which can be viewed as a measure of the ``Slaterness'' of the state
$\Psi$. A result of this form was first derived by Bach in \cite{Bach-92},
where he proved the following (see also Bach's paper \cite{Bach-93}
and Graf-Solovej's paper \cite{GrafSolovej-94} for related generalizations):
\begin{thm*}[Bach \cite{Bach-92}]
Let $X:\mathfrak{h}\rightarrow\mathfrak{h}$ be an orthogonal projection.
Then for any normalized $\Psi\in\bigwedge^{N}\mathfrak{h}$ it holds
that
\[
\mathrm{tr}\mleft(\mleft(X\otimes X\mright)\gamma_{2}^{\Psi,T}\mright)\geq-\,\mathrm{tr}\mleft(X\gamma_{1}^{\Psi}\mright)\cdot\min\left\{ 1,7.554\sqrt{\mathrm{tr}\mleft(X\mleft(\gamma_{1}^{\Psi}\mleft(1-\gamma_{1}^{\Psi}\mright)\mright)\mright)}\right\} .
\]
\end{thm*}
This can be seen as a sort of lower bound for $\gamma_{2}^{\Psi,T}$.
Below we will complement this result by proving the following Hilbert-Schmidt
bound for $\gamma_{2}^{\Psi,T}$:
\begin{thm}
\label{them:TruncatedEstimate}For any $N\in\mathbb{N}$ and $\Psi\in\bigwedge^{N}\mathfrak{h}$
it holds that
\[
\left\Vert \gamma_{2}^{\Psi,T}\right\Vert _{\mathrm{HS}}\leq\sqrt{5N\,\mathrm{tr}\mleft(\gamma_{1}^{\Psi}\mleft(1-\gamma_{1}^{\Psi}\mright)\mright)}.
\]
\end{thm}

We remark that at worst $\mathrm{tr}\mleft(\gamma_{1}^{\Psi}\mleft(1-\gamma_{1}^{\Psi}\mright)\mright)\leq\mathrm{tr}\mleft(\gamma_{1}^{\Psi}\mright)=N$,
in which case this bound essentially reduces to that of Theorem \ref{them:HilbertSchmidtBound}
(up to the quantity $\left\Vert \mleft(1-\mathrm{Ex}\mright)\gamma_{1}^{\Psi}\otimes\gamma_{1}^{\Psi}\right\Vert _{\mathrm{HS}}$),
but in view of that theorem this bound is non-trivial for any $\Psi$
with $\mathrm{tr}\mleft(\gamma_{1}^{\Psi}\mleft(1-\gamma_{1}^{\Psi}\mright)\mright)\ll N$.

\subsection{Acknowledgements}

I am grateful to Phan Th\`anh Nam for informing me of the application
of Theorem \ref{them:HilbertSchmidtBound} to the conjecture of Carlen-Lieb-Reuvers,
and for providing me with valuable feedback during the writing of
this paper.

This work was funded by the Deutsche Forschungsgemeinschaft (DFG,
German Research Foundation) -- Project-ID 470903074 -- TRR 352.

\section{Proof of the Theorems}

To prove Theorem \ref{them:HilbertSchmidtBound} we first observe
that by the characterization $\left\Vert T\right\Vert _{\mathrm{HS}}=\sup_{A}\left|\mathrm{tr}\mleft(AT\mright)\right|$,
where the supremum is over all Hilbert-Schmidt operators $A$ with
$\left\Vert A\right\Vert _{\mathrm{HS}}=1$, it suffices to estimate
a quantity of the form $\left|\mathrm{tr}\mleft(A\gamma_{2}^{\Psi}\mright)\right|$.
To do this we begin by writing
\begin{align}
\mathrm{tr}\mleft(A\gamma_{2}^{\Psi}\mright) & =\sum_{k,l,m,n=1}^{\infty}\left\langle \mleft(u_{k}\otimes u_{l}\mright),A\mleft(u_{m}\otimes u_{n}\mright)\right\rangle \left\langle \Psi,c_{k}^{\ast}c_{l}^{\ast}c_{n}c_{m}\Psi\right\rangle \\
 & =-\sum_{n=1}^{\infty}\left\langle \sum_{k,l,m=1}^{\infty}\overline{A_{k,l,m,n}}c_{m}^{\ast}c_{l}c_{k}\Psi,c_{n}\Psi\right\rangle \nonumber 
\end{align}
where $A_{k,l,m,n}=\left\langle \mleft(u_{k}\otimes u_{l}\mright),A\mleft(u_{m}\otimes u_{n}\mright)\right\rangle $.
The Cauchy-Schwarz inequality thus allows us to estimate
\begin{align}
\left|\mathrm{tr}\mleft(A\gamma_{2}^{\Psi}\mright)\right| & \leq\sum_{n=1}^{\infty}\left\Vert \sum_{k,l,m=1}^{\infty}\overline{A_{k,l,m,n}}c_{m}^{\ast}c_{l}c_{k}\Psi\right\Vert \left\Vert c_{n}\Psi\right\Vert \leq\sqrt{\sum_{n=1}^{\infty}\left\Vert \sum_{k,l,m=1}^{\infty}\overline{A_{k,l,m,n}}c_{m}^{\ast}c_{l}c_{k}\Psi\right\Vert ^{2}}\sqrt{\sum_{n=1}^{\infty}\left\Vert c_{n}\Psi\right\Vert ^{2}} \nonumber \\
 & \leq\sqrt{N\sum_{n=1}^{\infty}\left\langle \Psi,\left\{ \mleft(\sum_{k,l,m=1}^{\infty}\overline{A_{k,l,m,n}}c_{m}^{\ast}c_{l}c_{k}\mright)^{\ast},\sum_{k,l,m=1}^{\infty}\overline{A_{k,l,m,n}}c_{m}^{\ast}c_{l}c_{k}\right\} \Psi\right\rangle } \label{eq:MainTraceEstimate}
\end{align}
where we recognized that $\sum_{n=1}^{\infty}\left\Vert c_{n}\Psi\right\Vert ^{2}=\left\langle \Psi,\mathcal{N}\Psi\right\rangle =N$,
and used that as in the proof of the inequality $\gamma_{1}^{\Psi}\leq1$,
we are free to add the quantity $\left\Vert \mleft(\sum_{k,l,m=1}^{\infty}\overline{A_{k,l,m,n}}c_{m}^{\ast}c_{l}c_{k}\mright)^{\ast}\Psi\right\Vert ^{2}$
to obtain the anticommutator of $\sum_{k,l,m=1}^{\infty}\overline{A_{k,l,m,n}}c_{m}^{\ast}c_{l}c_{k}$
with its adjoint.

The reason for this last step is an argument which was recently used
in the study of the correlation energy of Fermi gases in \cite{ChrHaiNam-22},
although it has since been learned this was first used by Bell in
\cite{Bell-62}: Although the CAR does not imply a useful anticommutator
identity for sums of an \textit{even} number of creation and annihilation
operators, it does for sums of an \textit{odd} number of such operators.

In the present case we note the following identity:
\begin{prop}
Let $A_{k,l,m,n}\in\mathbb{C}$ for $k,l,m,n\in\mathbb{N}$ obey $A_{l,k,m,n}=-A_{k,l,m,n}$.
Then it holds for any $n\in\mathbb{N}$ that $T_{n}=\left\{ \mleft(\sum_{k,l,m=1}^{\infty}\overline{A_{k,l,m,n}}c_{m}^{\ast}c_{l}c_{k}\mright)^{\ast},\sum_{k,l,m=1}^{\infty}\overline{A_{k,l,m,n}}c_{m}^{\ast}c_{l}c_{k}\right\} $
is given by
\[
T_{n}=\sum_{m=1}^{\infty}\left|\sum_{k,l=1}^{\infty}\overline{A_{k,l,m,n}}c_{l}c_{k}\right|^{2}+4\sum_{k=1}^{\infty}\left|\sum_{l,m=1}^{\infty}A_{k,l,m,n}c_{l}^{\ast}c_{m}\right|^{2}-2\sum_{k,l=1}^{\infty}\left|\sum_{m=1}^{\infty}A_{k,l,m,n}c_{m}\right|^{2}.
\]
\end{prop}

\textbf{Proof:} By applying the CAR we find that for any $k,l,m,k',l',m'\in\mathbb{N}$
\begin{align}
\left\{ c_{k}^{\ast}c_{l}^{\ast}c_{m},c_{m'}^{\ast}c_{l'}c_{k'}\right\}  & =-c_{k}^{\ast}c_{l}^{\ast}c_{m'}^{\ast}c_{m}c_{l'}c_{k'}+\delta_{m,m'}c_{k}^{\ast}c_{l}^{\ast}c_{l'}c_{k'}+c_{m'}^{\ast}c_{l'}c_{k'}c_{k}^{\ast}c_{l}^{\ast}c_{m}\nonumber \\
 & =-c_{m'}^{\ast}c_{k}^{\ast}c_{l}^{\ast}c_{l'}c_{k'}c_{m}+\delta_{m,m'}c_{k}^{\ast}c_{l}^{\ast}c_{l'}c_{k'}+c_{m'}^{\ast}c_{l'}c_{k'}c_{k}^{\ast}c_{l}^{\ast}c_{m}\\
 & =-c_{m'}^{\ast}\left[c_{k}^{\ast}c_{l}^{\ast},c_{l'}c_{k'}\right]c_{m}+\delta_{m,m'}c_{k}^{\ast}c_{l}^{\ast}c_{l'}c_{k'},\nonumber 
\end{align}
and by standard commutator identities
\begin{align}
 & \quad\;\left[c_{k}^{\ast}c_{l}^{\ast},c_{l'}c_{k'}\right]=c_{k}^{\ast}\left[c_{l}^{\ast},c_{l'}c_{k'}\right]+\left[c_{k}^{\ast},c_{l'}c_{k'}\right]c_{l}^{\ast}\nonumber \\
 & =c_{k}^{\ast}\mleft(\left\{ c_{l}^{\ast},c_{l'}\right\} c_{k'}-c_{l'}\left\{ c_{l}^{\ast},c_{k'}\right\} \mright)+\mleft(\left\{ c_{k}^{\ast},c_{l'}\right\} c_{k'}-c_{l'}\left\{ c_{k}^{\ast},c_{k'}\right\} \mright)c_{l}^{\ast}\\
 & =\delta_{l,l'}c_{k}^{\ast}c_{k'}-\delta_{l,k'}c_{k}^{\ast}c_{l'}+\delta_{k,l'}c_{k'}c_{l}^{\ast}-\delta_{k,k'}c_{l'}c_{l}^{\ast}\nonumber \\
 & =-\delta_{k,k'}c_{l'}c_{l}^{\ast}-\delta_{l,l'}c_{k'}c_{k}^{\ast}+\delta_{k,l'}c_{k'}c_{l}^{\ast}+\delta_{l,k'}c_{l'}c_{k}^{\ast}+\delta_{k,k'}\delta_{l,l'}-\delta_{k,l'}\delta_{l,k'},\nonumber 
\end{align}
so
\begin{align}
\left\{ c_{k}^{\ast}c_{l}^{\ast}c_{m},c_{m'}^{\ast}c_{l'}c_{k'}\right\}  & =\delta_{m,m'}c_{k}^{\ast}c_{l}^{\ast}c_{l'}c_{k'}+\delta_{k,k'}c_{m'}^{\ast}c_{l'}c_{l}^{\ast}c_{m}+\delta_{l,l'}c_{m'}^{\ast}c_{k'}c_{k}^{\ast}c_{m}\\
 & -\delta_{k,l'}c_{m'}^{\ast}c_{k'}c_{l}^{\ast}c_{m}-\delta_{l,k'}c_{m'}^{\ast}c_{l'}c_{k}^{\ast}c_{m}-\delta_{k,k'}\delta_{l,l'}c_{m'}^{\ast}c_{m}+\delta_{k,l'}\delta_{l,k'}c_{m'}^{\ast}c_{m}.\nonumber 
\end{align}
Consequently $T_{n}$ is given by
\begin{align}
T_{n} & =\sum_{k,l,m=1}^{\infty}\sum_{k',l',m'=1}^{\infty}A_{k,l,m,n}\overline{A_{k',l',m',n}}\left\{ c_{k}^{\ast}c_{l}^{\ast}c_{m},c_{m'}^{\ast}c_{l'}c_{k'}\right\} \nonumber \\
 & =\sum_{m=1}^{\infty}\left|\sum_{k',l'=1}^{\infty}\overline{A_{k',l',m,n}}c_{l'}c_{k'}\right|^{2}+\sum_{k=1}^{\infty}\left|\sum_{l,m=1}^{\infty}A_{k,l,m,n}c_{l}^{\ast}c_{m}\right|^{2}+\sum_{l=1}^{\infty}\left|\sum_{k,m=1}^{\infty}A_{k,l,m,n}c_{k}^{\ast}c_{m}\right|^{2} \nonumber \\ 
 & -\sum_{k=1}^{\infty}\mleft(\sum_{k',m'=1}^{\infty}A_{k',k,m',n}c_{k'}^{\ast}c_{m'}\mright)^{\ast}\mleft(\sum_{l,m=1}^{\infty}A_{k,l,m,n}c_{l}^{\ast}c_{m}\mright) \\
 & -\sum_{l=1}^{\infty}\mleft(\sum_{l',m'=1}^{\infty}A_{l,l',m',n}c_{l'}^{\ast}c_{m'}\mright)^{\ast}\mleft(\sum_{k,m=1}^{\infty}A_{k,l,m,n}c_{k}^{\ast}c_{m}\mright)\nonumber \\
 & -\sum_{k,l=1}^{\infty}\left|\sum_{m=1}^{\infty}A_{k,l,m,n}c_{m}\right|^{2}+\sum_{k,l=1}^{\infty}\mleft(\sum_{m'=1}^{\infty}A_{l,k,m',n}c_{m'}\mright)^{\ast}\mleft(\sum_{m=1}^{\infty}A_{k,l,m,n}c_{m}\mright).\nonumber 
\end{align}
The claim now follows by renaming indices and using the assumed antisymmetry
of $A_{k,l,m,n}$.

$\hfill\square$

With this identity we now obtain the central estimate of this paper:
\begin{prop}
\label{prop:MainOperatorEstimate}For any Hilbert-Schmidt operator
$A:\mathfrak{h}\otimes\mathfrak{h}\rightarrow\mathfrak{h}\otimes\mathfrak{h}$
and $N\in\mathbb{N}$ it holds on $\bigwedge^{N}\mathfrak{h}$ that
\[
\sum_{n=1}^{\infty}\left\{ \mleft(\sum_{k,l,m=1}^{\infty}\overline{A_{k,l,m,n}}c_{m}^{\ast}c_{l}c_{k}\mright)^{\ast},\sum_{k,l,m=1}^{\infty}\overline{A_{k,l,m,n}}c_{m}^{\ast}c_{l}c_{k}\right\} \leq5N\left\Vert A\right\Vert _{\mathrm{HS}}^{2}.
\]
\end{prop}

\textbf{Proof:} We begin by noting that as
\begin{equation}
\sum_{k,l,m=1}^{\infty}\overline{A_{k,l,m,n}}c_{m}^{\ast}c_{l}c_{k}=\frac{1}{2}\sum_{k,l,m=1}^{\infty}\overline{A_{k,l,m,n}}c_{m}^{\ast}\mleft(c_{l}c_{k}-c_{k}c_{l}\mright)=\sum_{k,l,m=1}^{\infty}\frac{\overline{A_{k,l,m,n}-A_{l,k,m,n}}}{2}c_{m}^{\ast}c_{l}c_{k}
\end{equation}
and $\sum_{k,l,m,n}\left|A_{k,l,m,n}-A_{l,k,m,n}\right|^{2}\leq4\sum_{k,l,m,n}\left|A_{k,l,m,n}\right|^{2}$,
we can assume without loss of generality that $A_{l,k,m,n}=-A_{k,l,m,n}$.

It then follows from the previous proposition that
\begin{align}
T & :=\sum_{n=1}^{\infty}\left\{ \mleft(\sum_{k,l,m=1}^{\infty}\overline{A_{k,l,m,n}}c_{m}^{\ast}c_{l}c_{k}\mright)^{\ast},\sum_{k,l,m=1}^{\infty}\overline{A_{k,l,m,n}}c_{m}^{\ast}c_{l}c_{k}\right\} \\
 & \leq\sum_{m,n=1}^{\infty}\left|\sum_{k,l=1}^{\infty}\overline{A_{k,l,m,n}}c_{l}c_{k}\right|^{2}+4\sum_{k,n=1}^{\infty}\left|\sum_{l,m=1}^{\infty}A_{k,l,m,n}c_{l}^{\ast}c_{m}\right|^{2},\nonumber 
\end{align}
and the same argument we applied in equation (\ref{eq:2-BodyOperatorEstimate})
now shows that for any $\Psi\in\bigwedge^{N}\mathfrak{h}$
\begin{align}
 & \quad  \, \sum_{m,n=1}^{\infty}\left\Vert \sum_{k,l=1}^{\infty}\overline{A_{k,l,m,n}}c_{l}c_{k}\Psi\right\Vert ^{2} \leq\sum_{m,n=1}^{\infty}\mleft(\sum_{k=1}^{\infty}\left\Vert \sum_{l=1}^{\infty}\overline{A_{k,l,m,n}}c_{l}c_{k}\Psi\right\Vert \mright)^{2} \nonumber \\
 & \leq\sum_{m,n=1}^{\infty}\mleft(\sum_{k=1}^{\infty}\sqrt{\sum_{l=1}^{\infty}\left|A_{k,l,m,n}\right|^{2}}\left\Vert c_{k}\Psi\right\Vert \mright)^{2} \leq\mleft(\sum_{k,l,m,n=1}^{\infty}\left|A_{k,l,m,n}\right|^{2}\mright)\mleft(\sum_{k=1}^{\infty}\left\Vert c_{k}\Psi\right\Vert ^{2}\mright) \\
 & =N\left\Vert A\right\Vert _{\mathrm{HS}}^{2}\left\Vert \Psi\right\Vert ^{2}, \nonumber
\end{align}
i.e. $\sum_{m,n=1}^{\infty}\left|\sum_{k,l=1}^{\infty}\overline{A_{k,l,m,n}}c_{l}c_{k}\right|^{2}\leq N\left\Vert A\right\Vert _{\mathrm{HS}}^{2}$,
and as also $\left\Vert \sum_{k}\alpha_{k}c_{k}^{\ast}\right\Vert _{\mathrm{op}}\leq\sqrt{\sum_{k}\left|\alpha_{k}\right|^{2}}$,
it similarly holds that
\begin{equation}
\sum_{k,n=1}^{\infty}\left\Vert \sum_{l,m=1}^{\infty}A_{k,l,m,n}c_{l}^{\ast}c_{m}\Psi\right\Vert ^{2}\leq N\left\Vert A\right\Vert _{\mathrm{HS}}^{2}.
\end{equation}
$\hfill\square$

As noted at the beginning of this section, Theorem \ref{them:HilbertSchmidtBound}
immediately follows since we now have from equation (\ref{eq:MainTraceEstimate})
that
\begin{equation}
\left|\mathrm{tr}\mleft(A\gamma_{2}^{\Psi}\mright)\right|\leq\sqrt{5}N\left\Vert A\right\Vert _{\mathrm{HS}}.
\end{equation}

\subsubsection*{Proof of Theorem \ref{them:TruncatedEstimate}}

We note the following identity for $\gamma_{2}^{\Psi,T}$:
\begin{prop}
For any $\varphi_{1},\varphi_{2},\psi_{1},\psi_{2}\in\mathfrak{h}$
it holds that
\begin{align*}
\left\langle \mleft(\varphi_{1}\otimes\varphi_{2}\mright),\gamma_{2}^{\Psi,T}\mleft(\psi_{1}\otimes\psi_{2}\mright)\right\rangle  & =\left\langle \Psi,c\mleft(\gamma_{1}^{\Psi}\varphi_{2}\mright)c^{\ast}\mleft(\psi_{1}\mright)c^{\ast}\mleft(\psi_{2}\mright)c\mleft(\varphi_{1}\mright)\Psi\right\rangle \\
 & -\left\langle \Psi,c^{\ast}\mleft(\psi_{1}\mright)c^{\ast}\mleft(\psi_{2}\mright)c\mleft(\varphi_{1}\mright)c\mleft(\mleft(1-\gamma_{1}^{\Psi}\mright)\varphi_{2}\mright)\Psi\right\rangle .
\end{align*}
\end{prop}

\textbf{Proof:} By the CAR we have that
\begin{align}
c^{\ast}\mleft(\psi_{1}\mright)c^{\ast}\mleft(\psi_{2}\mright)c\mleft(\varphi_{2}\mright)c\mleft(\varphi_{1}\mright) & =-c^{\ast}\mleft(\psi_{1}\mright)c\mleft(\varphi_{2}\mright)c^{\ast}\mleft(\psi_{2}\mright)c\mleft(\varphi_{1}\mright)+\left\langle \varphi_{2},\psi_{2}\right\rangle c^{\ast}\mleft(\psi_{1}\mright)c\mleft(\varphi_{1}\mright)\nonumber \\
 & =c\mleft(\varphi_{2}\mright)c^{\ast}\mleft(\psi_{1}\mright)c^{\ast}\mleft(\psi_{2}\mright)c\mleft(\varphi_{1}\mright)\\
 & +\left\langle \varphi_{2},\psi_{2}\right\rangle c^{\ast}\mleft(\psi_{1}\mright)c\mleft(\varphi_{1}\mright)-\left\langle \varphi_{2},\psi_{1}\right\rangle c^{\ast}\mleft(\psi_{2}\mright)c\mleft(\varphi_{1}\mright),\nonumber 
\end{align}
which by applying $\left\langle \Psi,\mleft(\cdot\mright)\Psi\right\rangle $
yields
\begin{align}
\left\langle \mleft(\varphi_{1}\otimes\varphi_{2}\mright),\gamma_{2}^{\Psi}\mleft(\psi_{1}\otimes\psi_{2}\mright)\right\rangle  & =\left\langle \Psi,c\mleft(\varphi_{2}\mright)c^{\ast}\mleft(\psi_{1}\mright)c^{\ast}\mleft(\psi_{2}\mright)c\mleft(\varphi_{1}\mright)\Psi\right\rangle \\
 & +\left\langle \mleft(\varphi_{1}\otimes\varphi_{2}\mright),\mleft(\gamma_{1}^{\Psi}\otimes1\mright)\mleft(1-\mathrm{Ex}\mright)\mleft(\psi_{1}\otimes\psi_{2}\mright)\right\rangle .\nonumber 
\end{align}
Consequently
\begin{align}
 & \quad\,\left\langle \mleft(\varphi_{1}\otimes\varphi_{2}\mright),\gamma_{2}^{\Psi}\mleft(\psi_{1}\otimes\psi_{2}\mright)\right\rangle =\left\langle \mleft(\varphi_{1}\otimes\gamma_{1}^{\Psi}\varphi_{2}\mright),\gamma_{2}^{\Psi}\mleft(\psi_{1}\otimes\psi_{2}\mright)\right\rangle +\left\langle \mleft(\varphi_{1}\otimes\mleft(1-\gamma_{1}^{\Psi}\mright)\varphi_{2}\mright),\gamma_{2}^{\Psi}\mleft(\psi_{1}\otimes\psi_{2}\mright)\right\rangle \nonumber \\
 & =\left\langle \Psi,c\mleft(\gamma_{1}^{\Psi}\varphi_{2}\mright)c^{\ast}\mleft(\psi_{1}\mright)c^{\ast}\mleft(\psi_{2}\mright)c\mleft(\varphi_{1}\mright)\Psi\right\rangle -\left\langle \Psi,c^{\ast}\mleft(\psi_{1}\mright)c^{\ast}\mleft(\psi_{2}\mright)c\mleft(\varphi_{1}\mright)c\mleft(\mleft(1-\gamma_{1}^{\Psi}\mright)\varphi_{2}\mright)\Psi\right\rangle \\
 & +\left\langle \mleft(\varphi_{1}\otimes\varphi_{2}\mright),\mleft(1\otimes\gamma_{1}^{\Psi}\mright)\mleft(\gamma_{1}^{\Psi}\otimes1\mright)\mleft(1-\mathrm{Ex}\mright)\mleft(\psi_{1}\otimes\psi_{2}\mright)\right\rangle \nonumber 
\end{align}
which implies the claim since $\mleft(1\otimes\gamma_{1}^{\Psi}\mright)\mleft(\gamma_{1}^{\Psi}\otimes1\mright)\mleft(1-\mathrm{Ex}\mright)=\mleft(1-\mathrm{Ex}\mright)\mleft(\gamma_{1}^{\Psi}\otimes\gamma_{1}^{\Psi}\mright)$.

$\hfill\square$

Theorem \ref{them:TruncatedEstimate} can now be concluded in the
following form:
\begin{prop}
For any Hilbert-Schmidt operator $A:\mathfrak{h}\otimes\mathfrak{h}\rightarrow\mathfrak{h}\otimes\mathfrak{h}$,
$N\in\mathbb{N}$ and normalized $\Psi\in\bigwedge^{N}\mathfrak{h}$
it holds that
\[
\left|\mathrm{tr}\mleft(A\gamma_{2}^{\Psi,T}\mright)\right|\leq\sqrt{5N\,\mathrm{tr}\mleft(\gamma_{1}^{\Psi}\mleft(1-\gamma_{1}^{\Psi}\mright)\mright)}\left\Vert A\right\Vert _{\mathrm{HS}}.
\]
\end{prop}

\textbf{Proof:} By the previous proposition we can write the trace
as
\begin{align}
\mathrm{tr}\mleft(A\gamma_{2}^{\Psi,T}\mright) & =\sum_{k,l,m,n=1}^{\infty}\left\langle \mleft(u_{k}\otimes u_{l}\mright),A\mleft(u_{m}\otimes u_{n}\mright)\right\rangle \left\langle \mleft(u_{m}\otimes u_{n}\mright),\gamma_{2}^{\Psi,T}\mleft(u_{k}\otimes u_{l}\mright)\right\rangle \nonumber \\
 & =\sum_{n=1}^{\infty}\left\langle c^{\ast}\mleft(\gamma_{1}^{\Psi}u_{n}\mright)\Psi,\mleft(\sum_{k,l,m=1}^{\infty}\overline{A_{k,l,m,n}}c_{m}^{\ast}c_{l}c_{k}\mright)^{\ast}\Psi\right\rangle \\
 & -\sum_{n=1}^{\infty}\left\langle \mleft(\sum_{k,l,m=1}^{\infty}\overline{A_{k,l,m,n}}c_{m}^{\ast}c_{l}c_{k}\Psi\mright),c\mleft(\mleft(1-\gamma_{1}^{\Psi}\mright)u_{n}\mright)\Psi\right\rangle ,\nonumber 
\end{align}
so by Cauchy-Schwarz and Proposition \ref{prop:MainOperatorEstimate}
\begin{align}
\left|\mathrm{tr}\mleft(A\gamma_{2}^{\Psi,T}\mright)\right| & \leq\sqrt{\sum_{n=1}^{\infty}\left\langle \Psi,\sum_{n=1}^{\infty}\left\{ \mleft(\sum_{k,l,m=1}^{\infty}\overline{A_{k,l,m,n}}c_{m}^{\ast}c_{l}c_{k}\mright)^{\ast},\sum_{k,l,m=1}^{\infty}\overline{A_{k,l,m,n}}c_{m}^{\ast}c_{l}c_{k}\right\} \Psi\right\rangle }\nonumber \\
 & \qquad\qquad\qquad\qquad\cdot\sqrt{\sum_{n=1}^{\infty}\mleft(\left\Vert c^{\ast}\mleft(\gamma_{1}^{\Psi}u_{n}\mright)\Psi\right\Vert ^{2}+\left\Vert c\mleft(\mleft(1-\gamma_{1}^{\Psi}\mright)u_{n}\mright)\Psi\right\Vert ^{2}\mright)}\\
 & \leq\sqrt{5N\sum_{n=1}^{\infty}\mleft(\left\Vert c^{\ast}\mleft(\gamma_{1}^{\Psi}u_{n}\mright)\Psi\right\Vert ^{2}+\left\Vert c\mleft(\mleft(1-\gamma_{1}^{\Psi}\mright)u_{n}\mright)\Psi\right\Vert ^{2}\mright)}\left\Vert A\right\Vert _{\mathrm{HS}}.\nonumber 
\end{align}
Now we note that by the CAR
\begin{align}
\sum_{n=1}^{\infty}\left\Vert c^{\ast}\mleft(\gamma_{1}^{\Psi}u_{n}\mright)\Psi\right\Vert ^{2} & =\sum_{n=1}^{\infty}\left\langle \Psi,\mleft(\left\Vert \gamma_{1}^{\Psi}u_{n}\right\Vert ^{2}-c^{\ast}\mleft(\gamma_{1}^{\Psi}u_{n}\mright)c\mleft(\gamma_{1}^{\Psi}u_{n}\mright)\mright)\Psi\right\rangle \\
 & =\mathrm{tr}\mleft(\mleft(\gamma_{1}^{\Psi}\mright)^{2}-\mleft(\gamma_{1}^{\Psi}\mright)^{3}\mright)=\mathrm{tr}\mleft(\gamma_{1}^{\Psi}\mleft(\gamma_{1}^{\Psi}-\mleft(\gamma_{1}^{\Psi}\mright)^{2}\mright)\mright)\nonumber 
\end{align}
and
\begin{align}
\sum_{n=1}^{\infty}\left\Vert c\mleft(\mleft(1-\gamma_{1}^{\Psi}\mright)u_{n}\mright)\Psi\right\Vert ^{2} & =\sum_{n=1}^{\infty}\left\langle \Psi,c^{\ast}\mleft(\mleft(1-\gamma_{1}^{\Psi}\mright)u_{n}\mright)c\mleft(\mleft(1-\gamma_{1}^{\Psi}\mright)u_{n}\mright)\Psi\right\rangle \\
 & =\mathrm{tr}\mleft(\gamma_{1}^{\Psi}\mleft(1-\gamma_{1}^{\Psi}\mright)^{2}\mright)=\mathrm{tr}\mleft(\gamma_{1}^{\Psi}\mleft(1-2\gamma_{1}^{\Psi}+\mleft(\gamma_{1}^{\Psi}\mright)^{2}\mright)\mright)\nonumber 
\end{align}
for the claim.

$\hfill\square$

\section*{Data Availability Statement}
Data sharing not applicable to this article as no datasets were generated or analysed during the current study.

\end{document}